\documentclass{aa}
\usepackage{graphicx}
\def\gta{ \lower .75ex \hbox{$\sim$} \llap{\raise .27ex \hbox{$>$}} }
\def\lta{ \lower .75ex\hbox{$\sim$} \llap{\raise .27ex \hbox{$<$}} }
\begin{document}
   \title{High-z massive galaxies in the Hubble Deep Field South}

   \author{P. Saracco\inst{1}, M. Longhetti\inst{1}, E. Giallongo\inst{2},
S. Arnouts\inst{3}, S. Cristiani\inst{4}, S. D'Odorico\inst{5},  
A. Fontana\inst{2}, M. Nonino\inst{4}, E. Vanzella\inst{5,6}
          }

   \offprints{P. Saracco}

   \institute{INAF - Osservatorio Astronomico di Brera,
              Via Brera 28, 20121 Milano, Italy\\
\email{saracco, marcella, @brera.mi.astro.it}
         \and
           INAF - Osservatorio Astronomico di Roma, Via Frascati 33, 00040 Monte Porzio Catone, Italy  \\
             \email{giallo, fontana, @coma.mporzio.astro.it}
	\and
	Laboratoire d'Astronomie de Marseille, Traverse  du Siphon - BP 8, 13376 Marseille, France\\
	\email{stephane.arnouts@oamp.fr}
	\and
	INAF - Osservatorio Astronomico di Trieste, Via G. B. Tiepolo 11, 40131 Trieste, Italy\\
	\email{cristiani, nonino, @ts.astro.it }
\and
	European Southern Observatory, Karl-Schwarzschildstr. 2, D-85748 Garching, Germany \\
	\email{sdodoric, evanzell, @eso.org}
\and
Dipartimento di Astronomia dell'Universit\`a di Padova,
        Vicolo dell'Osservatorio 2,
        35122 Padova, Italy 
             }

   \date{Accepted}

   \abstract{A census of massive galaxies at redshift increasingly higher 
than $z\sim1$ may provide strong constraints on the history of mass assembly
and of star formation. 
Here we report the analysis of  three galaxies 
selected in the Hubble Deep Field South  at Ks$\le22$ on the basis of their unusually red
near-IR color J-K$\ge3$.
We have used population synthesis models to constrain their redshifts 
and their stellar masses.
One galaxy (HDFS-1269) is at redshift $z_{phot}\simeq2.4$ 
while the other two (HDFS-822 and HDFS-850) are at $z_{phot}\simeq2.9-3.0$.
All three galaxies have already assembled a stellar mass
of about $10^{11}$ M$_{\odot}$ at the observed redshift
placing the possible merging event of their formation at  z$\gta3.5$.
The inferred  mass weighted age of their stellar populations implies that
the bulk  of the stars formed at $z_f>3.5$.
The resulting co-moving density of $\mathcal{M}_{stars}\gta10^{11}$ M$_{\odot}$
galaxies at $\langle z\rangle\simeq2.7$ is
$\rho=1.2\pm0.7\times 10^{-4}$ Mpc$^{-3}$, about a factor two higher than the 
predictions of hierarchical models.
The comparison with the local density of galaxies implies that the three galaxies
must have already formed most of their stellar mass and that they cannot follow
an evolution significantly different from a passive aging.
The comparison with the density of local L$\ge$L$^*$ early types 
(passively evolved galaxies) suggests that their co-moving density 
cannot decrease by  more than a factor 2.5-3 from $z=0$ to $z\simeq3$ 
suggesting that up to 40\% of the stellar mass content of bright (L$\ge$L$^*$) 
local early type galaxies was already in place at $z>2.5$. 

   }

\titlerunning{High-z massive galaxies in the HDF-S}
\authorrunning{P. Saracco et al.}
   \maketitle
%

\section{Introduction}
Deep near-IR surveys are unveiling  
sources with  unusually red near-IR colors (J-K$>3$).
They are extremely rare at magnitudes brighter
than K=20 while their surface density  increases 
at fainter magnitudes.
Only one source redder than J-K=3  is present in the
the 65 arcmin$^2$ surveyed by Hall et al. (2001) down to K$\simeq$19.5
while 5 of them appear at Ks$\le21$ 
over a sub-area of 43 arcmin$^2$ of the ESO Imaging Survey
(Scodeggio and Silva 2000).
One red object  (HDFN-JD1) 
was  found  in the Hubble Deep Field North (HDFN; Dickinson et al. 2000). 
It has a magnitude K$\simeq22$ and  no  counterpart at wavelengths 
shorter than 1.2 $\mu$m.
Maihara et al. (2001) and Totani et al. (2001) noticed the presence 
of four sources at magnitudes K'$\gta21$ with color J-K$>3$ in the 
Subaru Deep Field (SDF).
Objects with these unusually red near-IR colors were also noticed 
in the HDF-S by Saracco et al. (2001) at K$>20.5$.    

The nature of these sources has not yet been firmly
established even if it is quite certain  that they are not galactic objects.
Indeed, very low mass stars, such as L-dwarfs, can display colors 
only slightly redder  than J-K=2 (Chabrier et al. 2000; Kirkpatrick et al. 
2000). 
Stars heavily reddened by circumstellar dust due to undergoing mass loss, 
such as Mira variables and carbon stars, can be redder than L-dwarfs.
However, Whitelock et al. (1995; 2000) found only 2 stars having  J-K$\simeq3$ 
out of the 350 Mira and mass-losing stars observed.
It seems unlikely that the extremely small 
fields of the HDFs and of the SDF can contain so many extremely rare stars
at high galactic latitude.
Furthermore, the apparent K-band magnitude of these unusually red
objects (K$\gta20$) would place them out of the Galaxy at a distance 
larger than 5 Mpc if they were stars.

Cutri et al. (2001) and Smith et al. (2002) find 4 QSOs with colors 
J-K$>3$ among the 231 red AGNs selected using a J-K$>2$  criterion
from the Two Micron All-Sky Survey (2MASS).
They are brighter than K=14 and are at $z<0.3$. 
If the unusually red sources seen at K$\gta20$ were 
dominated by AGNs at these redshifts they would be at least 
10$^3$ times less luminous then the 2MASS AGNs, i.e. too faint to be AGNs.
{ On the other hand,  un-obscured AGNs at larger $z$ would get 
rapidly bluer since the rest-frame near-IR  excess 
would be redshifted beyond the K-band.
In the case of dust obscured AGNs, they should be reddened by at least
A$_V\simeq3$ mag and placed at $z\ge2$  to match the observed J-K color.
However, such values of extinction characterize AGNs for which the rest-frame 
optical luminosities are usually dominated by the continuum of the host 
galaxy (e.g. Maiolino et al. 2000).} 

The Extremely Red Objects (EROs) studied so far (e.g. Thompson et al. 1999; 
Cimatti et al. 1999, 2002; Daddi et al. 2000; McCarthy et al. 2001; 
Martini et al. 2001; Mannucci et al. 2002; Miyazaki et al. 2002)  
are characterized by  near-IR colors usually bluer than J-K$\simeq2.5$.
Colors redder than J-K$\sim2.5$ are not expected even for 
passively evolved galaxies down to $z\sim2$ 
(e.g. Saracco et al. 1999).
Indeed, all the  EROs spectroscopically observed so far 
lie  at $z<2$ (e.g. Cimatti et al. 2002; Saracco et al. 2003).
Thus, the unusually red near-IR color characterizing these objects
suggests redshifts $z\gta2$ and, possibly, a component of dust absorption.
Dickinson et al. (2000) consider various hypothesis for the nature
of HDFN-JD1: from the most extreme of an objects at $z\gta10$, justified
by the non-detection of the object from 0.3 to 1.1 $\mu$m, to the least
extreme of a dusty galaxy at $z>2$ or a maximally old
elliptical galaxy at $z\gta3$. 
The analysis of Hall et al. (2001)
suggests a redshift $z\sim2.4$ for their unusually red object.
Totani et al. (2001), by comparing the J-K color and the surface number density
of the red sources in the SDF with model predictions,
{ conclude that they are best explained
by dusty elliptical galaxies at $z\sim3$ in the starburst phase of their formation. 
Thus, the analysis of unusually red near-IR objects performed so far
place these galaxies at $z>2-3$.}
Our knowledge of the Universe at these redshifts comes mostly
from the Lyman-Break Galaxies (LBGs) selected through
the U-dropout method based on UV-optical color (e.g. Steidel et al. 1996).
The unusually red objects above are missed by this  
selection technique  both due to their faintness at optical wavelength 
and to their different optical colors (see e.g. Vanzella et al. 2001).
Consequently, also the information they bring relevant to the Universe at that $z$
are missed.
For this reason they could be extremely important to probe the Universe at 
these redshifts.

{ In this paper we present the analysis based on a multi-band data set
(from 0.3 $\mu$m to 2.15 $\mu$m) of  three J-K$\ge3$ sources
selected at Ks$\le22$ on the HDF-S.
The near-IR data have been collected by the Faint Infra-Red Extra-galactic 
Survey (FIRES, Franx et al. 2000).}
In \S 2 we present the imaging, the photometry and the spatial extent 
analysis of the three sources.
In \S 3 we derive  the redshift and, consequently, the stellar masses of 
the galaxies through the comparison of the data with population synthesis models.
In \S 4 we derive the co-moving spatial density of these objects
and we try to constrain their formation and evolution in \S 5.
We summarize our results in \S 6.
Throughout this paper, magnitudes are expressed in the Vega system 
unless explicitly stated otherwise.
We adopt an  $\Omega_m=0.3$, $\Omega_\Lambda=0.7$ cosmology with 
H$_0$=70 km s$^{-1}$  Mpc$^{-1}$.


\section{Imaging and photometric analysis}
The optical images  are the Version 2 of the 
Hubble Space Telescope images in the F300W, F450W, F606W and F814W 
filters (U$_{300}$, B$_{450}$, V$_{606}$ and I$_{814}$ hereafter, 
Casertano et al. 2000). 
{ Near-IR images centered in the HDF-S come from the FIRES project
(Franx et al. 2000) and they have been obtained with the ISAAC spectro-imager
at VLT-Antu  in the filters Js, H and Ks.
The three sources analyzed here were originally noticed by 
Saracco et al. (2001) in the analysis of a first set of such data 
($\sim8$ hours of exposure)} and turned out to be EROs (I-K$>4$) 
from the analysis of the multi-band data by 
Vanzella et al. (2001; object IDs: 822, 850 and 1269 hereafter).
The photometry and the analysis presented here are based on the final 
FIRES data set  (Labbe\'\ et al. 2003) summing up to about $\sim30$ hours exposure 
per filter with a measured FWHM$\sim0.55$ arcsec.
{ The near-IR data  have been reduced following the recipe
described in Saracco et al. (2001).} 
Photometric redshifts and analysis of the whole multi-band data 
sample is presented  in Fontana et al. (2003a).
In Fig. 1, the optical (B$_{450}$, V$_{606}$, I$_{814}$) and the near-IR
(Js, H and Ks) images of the three sources are shown.
The intensity levels in each image are constrained within 
-1$\sigma$ and 3$\sigma$ from the background.
Colors have been measured within the Ks-band detection isophote
by using Sextractor (Bertin \& Arnouts 1996) in double image mode.
{ To this end, we first rebinned the IR images 
to the same pixel size and orientation of the WFPC2 images.
Then, the WFPC2 images were smoothed to the same effective PSF of
the IR images as described in Vanzella et al. (2001).
Photometric errors have been obtained by rescaling the formal
SExtractor errors by a correction factor to take into account the
correlation of neighboring pixels. 
The procedure used to estimate the correction
factor is described in Vanzella et al. (2001).}
None of the three sources has a detectable flux in the U$_{300}$-band,
while only one of them (HDFS-1269) has a reliable flux in the B$_{450}$ band.
The photometry we obtained is in good agreement with the one
derived by Labbe\'\ et al. (2003).
In Tab. 1 the optical and near-IR colors of the three sources
are reported. 
It is worth  noting that  red AGNs have optical-IR
colors 1.5-2 mag bluer than those of our sources (Francis et al. 2000; 
Smith et al. 2002).
Thus, even if the presence of an AGN cannot be excluded 
(see e.g. van Dokkum et al. 2003), the optical-to-NIR
colors exclude, once more, that the continuum emission is dominated 
by nuclear activity.
 \begin{table*}
\caption{Photometry of the three EROs 1269, 822 and 850.
{ Magnitudes are in the Vega system. Colors are measured within the 
aperture defined by the Ks-band isophote. 
The Ks-band magnitude is the MAG\_AUTO measured by Sextractor.}}
\centerline{
\begin{tabular}{lccccccccc}
\hline
\hline
  Object  & RA(J2000)&Dec(J2000)& U$_{300}$-Ks& B$_{450}$-Ks&V$_{606}$-Ks& I$_{814}$-Ks& Js-Ks& H-Ks& Ks\\
  \hline
HDFS\_1269&22:32:49.25&-60:32:11.62& $>$7.7& 7.6$\pm0.5$& 6.9$\pm$0.2&5.5$\pm$0.1 &3.0$\pm$0.2&1.2$\pm$0.1&20.80$\pm$0.05\\
HDFS\_822& 22:32:53.39&-60:31:54.33& $>$7.6& $>$7.6&      7.0$\pm$0.3&5.7$\pm$0.2 &3.5$\pm$0.3&1.4$\pm$0.2&20.9$\pm$0.1\\
HDFS\_850& 22:32:53.01&-60:33:57.03& $>$6.8& $>$6.8&      7.1$\pm$0.3&5.6$\pm$0.2 &3.7$\pm$0.4&1.1$\pm$0.2&21.7$\pm$0.1\\
\hline
\hline
\end{tabular}
}
\end{table*}

   \begin{figure*}
   \centering
   \includegraphics{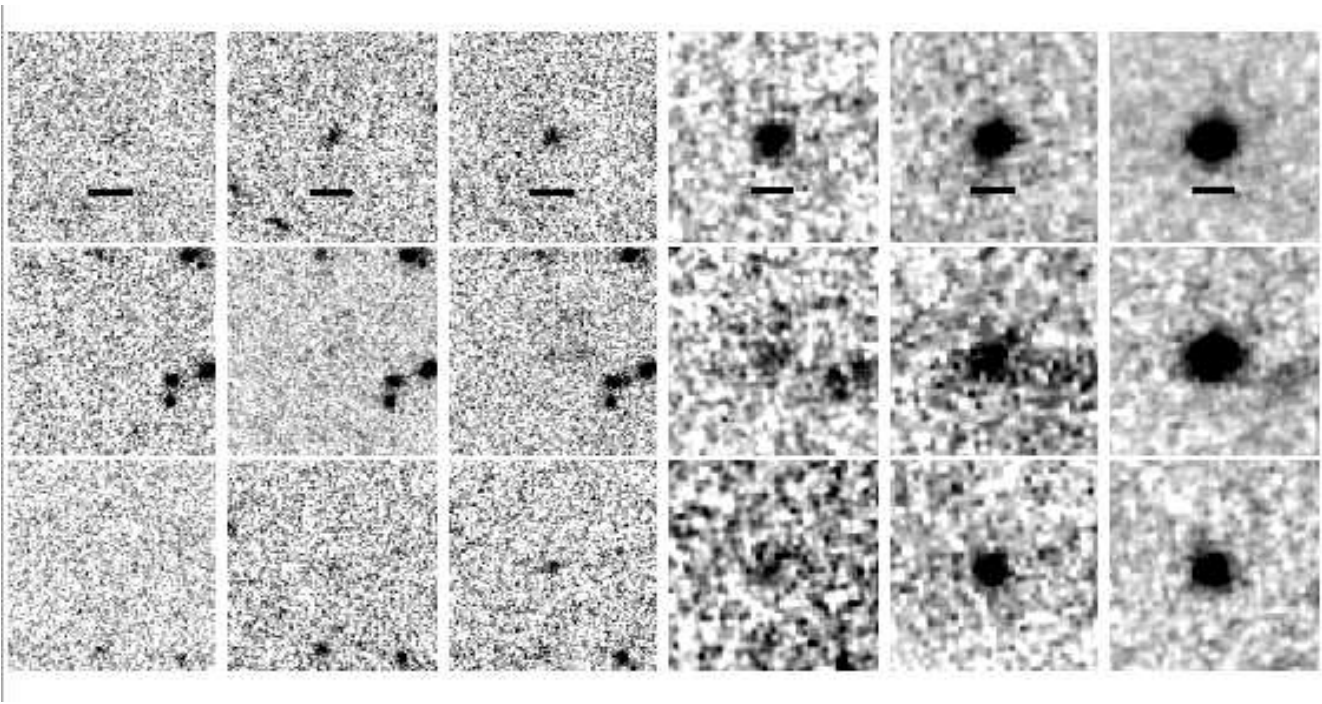}
\caption{
From left to right: B$_{450}$, V$_{606}$, I$_{814}$, 
Js, H and Ks band images of the three EROs 1269 (upper panel), 822 (middle 
panel) and 850 (lower panel). The images are 5$\times5$ arcsec centered
on the sources.  The intensity of each image is  in the range 
[-1$\sigma$;+3$\sigma$] from the  background. The lines shown
for reference in the images of 1269 are 1 arcsec width.} 
              \label{FigGam}%
    \end{figure*}
The three galaxies appear very compact in the K-band image.
In Fig. 2 we compare the K-band surface brightness profile
of the three galaxies  with the  profile of a bright star in the field.
The two brightest galaxies (HDFS-1269 and HDFS-822) are clearly
resolved  while HDFS-850 is not resolved.
Assuming that it is at a redshift comparable to the other two sources,
as suggested by its colors, it is $\sim$1 mag intrinsically fainter.
Using the magnitude-size relations in Snigula et al. (2002)
we found that, for the same profile, HDFS-850 would have a scale radius
$\sim$1.6 and $\sim$3 times narrower than  HDFS-822 (or HDFS-1269) 
in the case of an exponential and a de Vaucouleur profile respectively.
Thus, considering the angular extent of the two brightest galaxies, it is expected 
that the profile of HDFS-850 is dominated by the seeing.  

   \begin{figure}
   \centering
   \includegraphics[width=9cm]{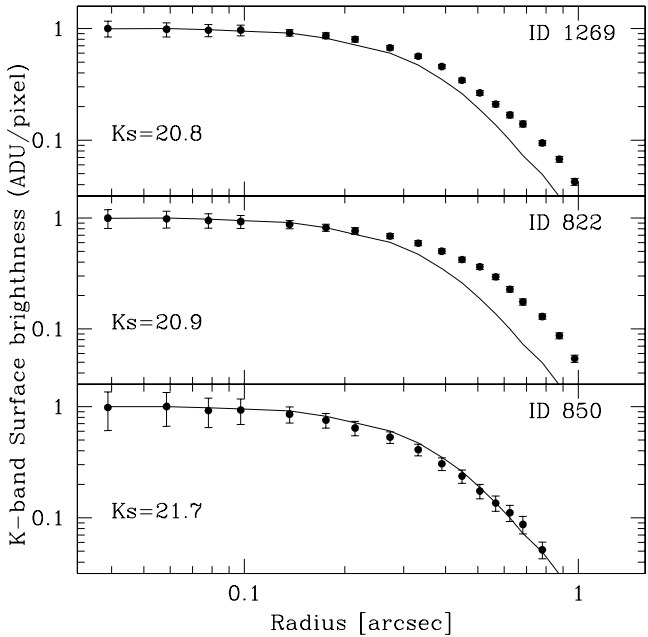}
\caption{Radial surface brightness profile of the three EROs discussed 
here (points) compared to one isolated star in the HDF-S (line). 
The profiles are normalized to the flux within the first aperture.} 
              \label{FigGam}%
    \end{figure}

\section{Photometric redshifts and  stellar masses of J-K$\ge$3 galaxies}
In this section we compare the data with population synthesis models
to constrain  the redshift and the stellar mass of the three galaxies. 
The grid of templates used is based on the 
latest version of the Bruzual \& Charlot (1993) models and it is
similar to the one  described in  Fontana et al. (2003a,b) in terms of
star formation histories (SFHs), metallicity and extinction (A$_V$).
{ The SFHs considered, besides the simple stellar population (SSP) and the 
constant star formation rate ({\em cst}), are described by an exponentially 
declining star formation rate (SFR) with  e-folding time ($\tau$) 
in the range 0.1-15 Gyr.}
In this paper, besides  the set of 
synthetic templates obtained with the Salpeter (Sal; Salpeter 1955)
 initial mass function  (IMF; 0.1M$_\odot<\mathcal{M}<100$M$_\odot$), 
we considered also three analogous sets generated with the 
Miller-Scalo (MS; Miller \& Scalo 1979), Scalo (Sca; Scalo 1986) and
Kroupa (Kro; Kroupa 2001) IMF respectively and a set based on the Kro IMF 
derived from  the spectrophotometric models of Charlot \& Longhetti (2001). 
{ These latter take into account consistently  the emission from stars  
and from the gas (Kro+em) and allow to probe the possibility that the 
extreme near-IR colors are due to emission lines rather than to  continuum 
emission (see e.g. Totani et al. 2001 and  Franx et al. 2003).}
The parameters used to generate the grid of synthetic templates are
summarized in Tab. 2.

The $\chi^2$-minimization procedure of {\em hyperz} (Bolzonella et al. 2000)
has been applied to find the best fitting spectral template to the 
observed colors for each set of IMF, SFH and  metallicity.
{ In practice, for a given IMF, metallicity and  SFH, we find the best fitting 
template, i.e. the corresponding redshift and  age.
In the best fitting procedure the extinction has been allowed to vary 
within the range $0\le A_V\le2$ mag} and, at each $z$, galaxies have been forced 
to have age lower than the Hubble time at that $z$. 
We repeated this procedure for the 5 IMFs, the 3 metallicity values 
and the 10 SFHs considered.
Among the 150 best fitting templates thus obtained we considered those
 having a probability P($\tilde\chi^2$)$>0.68$ (where $\tilde\chi^2$ is the reduced 
$\chi^2$).
For each galaxy, this set of templates has been used to define the range of 
variability of the relevant parameters we derived: $z$, SFHs and relevant age, 
stellar mass.   

\begin{table}
\caption{Parameters used to construct the grid of templates}
\centerline{
\begin{tabular}{ll}
\hline
SFH $\tau$ [Gyr] &  0.1, 0.3, 0.6, 1, 2, 3, 5, 15, SSP, {\em cst}\\
\hline
Metallicity & 0.2 $Z_\odot$, 0.4 $Z_\odot$, $Z_\odot$\\
\hline
A\_V [mag] & 0$\div$2 \\
\hline
Extinction law & Calzetti et al. (2000)\\
\hline
IMFs & Sal, Sca, MS, Kro, Kro+em\\
\hline
\end{tabular}
}
\end{table}

{ For each template considered, we derived the stellar mass 
$\mathcal{M}_{stars}$ of the galaxy from the mass-to-Ks-band light ratio
since it is relatively insensitive to the star formation history 
with respect to the optical bands (Charlot 1996).
The mass in this ratio (and thus $\mathcal{M}_{stars}$) results, 
in fact, by the integral of the SFR over 
the SFH of the template in the interval  $0\le t\le t_{age}$, where $t_{age}$ 
is the age of the template.
This mass corresponds to the processed gas involved in the star forming process
in this interval (see e.g. Madau et al. 1998). 
Thus, this is the mass that has been burned into stars at some time in this
interval, i.e. the mass that was contained and/or is still locked into stars.}
We have also considered the mass locked into stars ($\mathcal{M}_{*}$) 
at the age of the galaxy as resulting from the difference of $\mathcal{M}_{stars}$ 
and the processed gas returned to the interstellar medium through stellar winds.
As far as our three galaxies, we have found that
$\mathcal{M}_{*}\ge0.8\mathcal{M}_{stars}$ for Sal IMF and
$\mathcal{M}_{*}\ge0.6\mathcal{M}_{stars}$ for MS IMF, this latter being
the lowest fraction of mass locked into stars for the same value of
$\mathcal{M}_{stars}$.
The mass $\mathcal{M}_{*}$, is strongly related to the different mixture of stars 
and to their evolutionary stage.
Thus, it is much more dependent  on the IMF, the metallicity, 
the SFH and the extinction of the best fitting template with respect 
to $\mathcal{M}_{stars}$.
{ For these reasons, we adopted $\mathcal{M}_{stars}$ as the stellar mass 
of the galaxies since it is a robust estimate.}

{ The results of the best fitting procedure relevant to the 
Sal, MS and Kro IMFs are summarized in Tab. 3.
We do not report the results obtained with the Sca IMF since the best fits 
obtained with this IMF were always worst than those obtained with the other
IMFs and the values of the fitted parameters were within the ranges defined 
by the other models.   
The results obtained with the set of templates 
including emission lines (Kro+em) are not listed in the table since the
best fitting templates were always the same as those without emission lines (Kro).
This latter result shows that the observed extremely red near-IR colors are not 
likely to be dominated by emission lines.
In Tab. 4, we report for each galaxy the range of 
variability of the stellar mass ($\Delta\mathcal{M}_{stars}^{IMF}$) 
and of the mass-to-Ks-band light ratio ($\Delta$M/L$_K^{IMF}$) 
defined by the fitting models with different IMF. 

In the following discussion of the individual objects,
we compare the properties of local galaxies with those of our
three galaxies.
To this end,  we derived a lower limit to their Ks-band luminosities at $z=0$ 
applying an evolutionary correction factor.  
The adopted correction factor for each galaxy is an 
upper limit to the expected luminosity evolution in the relevant range of redshift
and corresponds to the passive aging of the youngest stellar population that 
could populate the galaxy at the relevant redshift.
{ The age  considered for the stellar population  is the mass weighted age 
described in \S 5.1. The youngest age of each galaxy is chosen  among the 
relevant best fitting templates (P($\tilde\chi^2>0.68$)). 
The passive aging is then traced from $z_{phot}$ to $z=0$ assuming a SSP
with age at $z_{phot}$ equal to the youngest (mass weighted) age.
It is worth noting that, for the relevant range of redshift, the expected luminosity
evolution is not significantly dependent on the assumed IMF (differences less than 
0.2 mag).}
We considered the Ks-band  to minimize the uncertainties in this extrapolation.
Indeed, the evolutionary corrections in the Ks-band are much smaller and 
less dependent on the SFH than those in the optical bands.
For instance, the evolutionary correction for a passive aging from $z=3$ to $z=0$ 
in the case of a SSP is $\sim1.7$ mag in Ks while it is $\sim2.8$ mag in V.}

\begin{table*}
\caption{Parameters resulting from the best fits of stellar population models
to the photometric data for Sal, MS and Kro IMFs. The $\tilde\chi^2$ represent the 
reduced $\chi^2$. 
The model which formally provides the best fit (the lowest $\tilde\chi^2$) 
is marked with ``*''. The V-band absolute magnitude has been derived by the Ks-band
apparent magnitude. {\bf The absolute magnitudes in the Ks and in the V bands
are corrected for the extinction.}}
\centerline{
\begin{tabular}{llllcccccccc}
\hline
\hline
ID & SFH & $\tilde\chi^2$& P($\tilde\chi^2$)& $z_{phot}$ & Age & A$_V$ & Z &M$_V$& M$_K$ & $^a$M/L$_K$&$\mathcal{M}_{stars}$\\
& &       &     &     & [Gyr] & [mag] & [Z$_\odot$]& [mag] &[mag]& [M/L]$_\odot$& [$10^{11}$M$_\odot$] \\ 
\hline
& Sal IMF  & & & & & & & & & &\\
HDFS\_1269$^*$&$\tau=0.3$   &0.33& 0.92& 2.4$^{+0.3}_{-0.2}$& 2.0 & 0.3  & 0.4&-22.9& -25.1& 0.50& 1.2\\
HDFS\_822     &$\tau=0.3$   &0.06& 0.99& 3.0$^{+0.3}_{-0.2}$& 1.7 & 1.0  & 0.2&-24.1& -26.1& 0.45& 3.3\\
HDFS\_850     &$\tau=0.1$   &0.16& 0.98& 2.9$^{+0.2}_{-0.3}$& 0.5 & 1.9  & 1.0&-24.1& -25.8& 0.30& 1.4\\
\hline
& MS IMF  & & & & & & & & & &\\
HDFS\_1269    &$\tau=0.3$   &0.34& 0.92& 2.5$^{+0.3}_{-0.2}$& 2.0 & 0.2  & 0.4&-22.9& -25.1& 0.26& 0.6\\
HDFS\_822$^*$ &$\tau=0.3$   &0.04& 0.99& 3.0$^{+0.3}_{-0.2}$& 1.7 & 0.9  & 0.2&-24.1& -26.1& 0.23& 1.4\\
HDFS\_850     &$\tau=0.1$   &0.17& 0.98& 2.9$^{+0.2}_{-0.3}$& 0.5 & 1.8  & 1.0&-24.0& -25.8& 0.14& 0.6\\
\hline
& Kro IMF  & & & & & & & & & &\\
HDFS\_1269    &$\tau=0.3$   &0.38& 0.89& 2.4$^{+0.3}_{-0.2}$& 2.0 & 0.4  & 0.4&-22.9& -25.1& 0.38& 0.9\\
HDFS\_822     &$\tau=0.3$   &0.04& 0.99& 3.0$^{+0.3}_{-0.2}$& 1.7 & 1.0  & 0.2&-24.1& -26.0& 0.35& 2.1\\
HDFS\_850$^*$ &$\tau=0.1$   &0.16& 0.98& 2.9$^{+0.2}_{-0.3}$& 0.5 & 1.9  & 1.0&-24.0& -25.8& 0.22& 1.0\\
\hline
\hline
\end{tabular}
}
{\hskip 1truecm $^a$ We used M$_{\odot,K}$=3.4 (Allen 1973)}
\end{table*}

   \begin{figure}
   \centering
   \includegraphics[width=9.5cm,height=8.5cm]{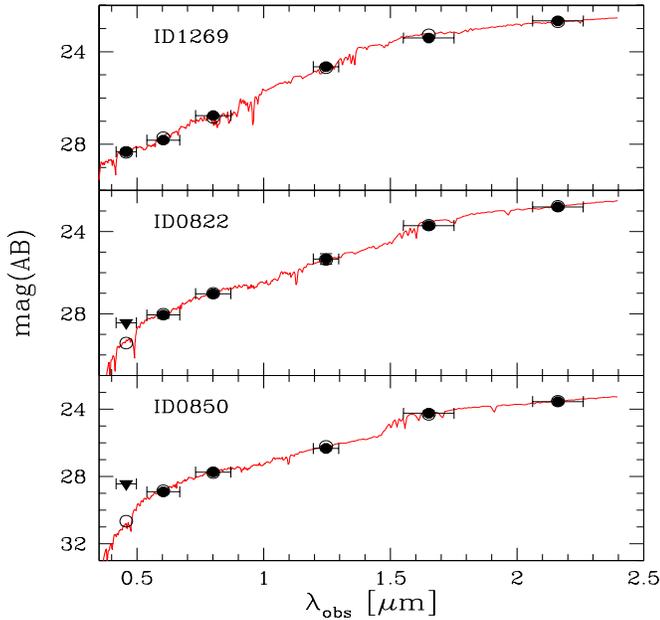}
\caption{The best fitting templates 
(see Tab. 3) are superimposed on the observed  photometric data 
(filled points) in the 
B$_{450}$, V$_{606}$, I$_{814}$, Js, H and Ks 
bands of the ERO 1269 (upper panel), 822 (middle panel) and 850
(lower panel) respectively. 
Magnitudes are in the AB system.
Filled triangles represent upper limits. Open circles are the magnitudes derived
from the best fitting template.
} 
              \label{FigGam}%
    \end{figure}
      
\paragraph{HDFS\_1269} -
This is the sole object for which Sextractor has reliably detected flux 
in the B$_{450}$-band. 
The best-fitting value to the redshift of  HDFS-1269  is
$z_{phot}=2.4^{+0.3}_{-0.2}$.
The quoted errors  are the formal uncertainties of the fitting procedure
(68\% confidence level).
The  rest-frame (k-corrected) absolute magnitude of this galaxy,
as resulting from its distant modulus (46.44) is M$_K\simeq-25.1$,
i.e. L$>$3L$^*$, where we considered M$^*_K\simeq-24.2$ for local galaxies 
(Cole et al. 2001; Huang et al. 2002).
The best fit is formally given by a 2 Gyr old, $\tau=0.3$ Gyr model
with an extinction A$_V=$0.3 and Z=0.4Z$_\odot$ obtained with a Sal IMF
(see Tab. 3).
In Fig. 3 (upper panel) the best-fitting template is shown together with
the photometric data points.
For this galaxy 61 exponentially $\tau\ge0.3$ Gyr decaying 
models  fit the data  ($0\le A_V\le2$), 
providing Age in the range 1-4 Gyr  and redshifts $1.9<z<2.7$.
{ The fitting models provide stellar masses of the order of
$10^{11}$M$_\odot$ independently on the IMF assumed, as shown in Tab. 4.
The stellar mass derived by the best fitting templates with the Sal IMF is 
 $\mathcal{M}_{stars}=1.2\times10^{11}$M$_\odot$.

The degeneracy with respect to the SFH, 
does not help in tracing the evolution of this galaxy
from $z_{phot}=2.4$ to $z=0$. 
Indeed, the stellar mass 
already assembled at $z=2.4$,  may represent 
from 35\% (if $\tau=15$ Gyr) to almost 100\% (if $\tau=0.3$ Gyr) of the stellar 
mass the galaxy could form down to $z=0$, depending on the SFH considered among the 61 models.}
The lower limit to the Ks-band luminosity that this galaxy
would have at $z=0$ is  L$_{z=0}\ge$L$_*$, having estimated a passive evolutionary
dimming  of $\Delta M_K\simeq 1.3$ mag  from $z_{phot}=2.4$ to $z=0$.

\paragraph{HDFS\_822} - 
The best-fitting value to the redshift  is
$z_{phot}=3.0^{+0.3}_{-0.2}$ and the resulting absolute magnitude
is M$_K\simeq-26.0$ (L$>$5L$^*$).
The best fit is formally given by a 1.7 Gyr old, $\tau=0.3$ Gyr model
with an extinction A$_V=$0.9 and Z=0.2Z$_\odot$ obtained with a MS IMF
and it  is shown in Fig. 3 (middle panel).
For this galaxy, 93 templates fit the data (0.4$\leq$A$_V\leq$2)
providing Age in the range 0.2-2.5 Gyr  
and redshifts 2.4$<z<3.2$
{ The stellar mass derived in the case of Sal IMF is 
 $\mathcal{M}_{stars}=3.3\times10^{11}$M$_\odot$.
For this galaxy the models provide a stellar mass  always 
larger than $10^{11}$M$_\odot$ independently on the IMF assumed.}
Also in this case, the degeneracy  with respect to the SFH implies that
 the stellar mass already formed at $z\simeq3$  may represent 
from 20\% to $\sim100$\% of the stellar mass the galaxy could have at $z$=0. 
The lower limit to the Ks-band luminosity of this galaxy at $z=0$
is L$_{z=0}>$1.2L$_*$ ($\Delta M_K\simeq1.7$ mag).

\paragraph{HDFS\_850} -
This is the faintest and the reddest of the three galaxies selected.
In this case the best-fitting template is given by 
a 0.5 Gyr old, $\tau=0.1$ Gyr model
with an extinction A$_V=$1.9 and Z=Z$_\odot$ obtained with a Kro IMF.
This model provides a redshift $z_{phot}=2.9^{+0.2}_{-0.3}$.
For this ERO, 58 templates (all the $\tau\le2$ Gyr models and SSP model) fit well  
the data with extinction 1$\leq$A$_V\leq$2, Age in the range 0.25-1.7 Gyr  and 
redshifts 2.6$<z<3.1$.
The stellar mass derived is $\mathcal{M}_{stars}=1.4\times10^{11}$M$_\odot$
(Sal IMF).
The stellar mass already formed and assembled
at $z\simeq2.9$  may represent from $\sim70$\% to $\sim100$\% of the stellar mass 
the galaxy could form down to $z$=0. 
The lower limit we derived to the Ks-band luminosity of this galaxy
 is L$_{z=0}\simeq$L$_*$ ($\Delta M_K\simeq1.5$ mag).

\begin{table*}
\caption{{ Ranges of stellar masses  ($\Delta\mathcal{M}_{stars}^{IMF}$)
and of mass-to-Ks-band light ratios ($\Delta$M/L$_K^{IMF}$) 
defined by the fitting models with different IMF (Sal, MS and Kro). 
}}
\centerline{
\begin{tabular}{lcccccc}
\hline
\hline
ID & $\Delta\mathcal{M}_{stars}^{Sal}$ &$\Delta$M/L$_K^{Sal}$& $\Delta\mathcal{M}_{stars}^{MS}$ &$\Delta$M/L$_K^{MS}$&$\Delta\mathcal{M}_{stars}^{Kro}$ &$\Delta$M/L$_K^{Kro}$\\
& [$10^{11}$M$_\odot$] &[M/L]$_\odot$ &  [$10^{11}$M$_\odot$] &[M/L]$_\odot$ & [$10^{11}$M$_\odot$] &[M/L]$_\odot$ \\ 
\hline
HDFS\_1269    &1.2-2.1&0.35-0.75 & 0.6-1.2&0.20-0.44 &0.9-1.5&0.29-0.49 \\
HDFS\_822     &2.3-5.6&0.25-0.52 & 1.1-2.6&0.14-0.27 &1.7-4.2&0.2-0.47 \\
HDFS\_850     &1.0-1.4&0.25-0.45 & 0.5-1.2&0.12-0.24 &0.8-1.1&0.18-0.28 \\
\hline
\hline
\end{tabular}
}
\end{table*}

\paragraph{} Extremely young  ages with active star formation and no dust 
attenuation are included among the best fitting templates. 
Hence, the minimum permissible mass-to-light ratio for each galaxy 
has been derived with the procedure adopted.
Consequently, the lower bound of the range of variability of the stellar mass
given in Tab. 4 should be considered a lower limit to the mass in stars of each
galaxy.
Moreover, active star formation can partially mask the presence of a more massive and
much older stellar population (see e.g. Papovich et al. 2001)
which could not be correctly accounted for using a single SFH as adopted 
in the present analysis.
The results we obtained show that {\em i}) all three galaxies are at redshift 
$2<z<3$; 
{\em ii}) they have a high stellar mass content  
($\mathcal{M}_{stars}\simeq10^{11}$ M$_\odot$) already formed and assembled at 
the observed redshift; {\em iii}) they would populate the  bright end 
(L$_{z=0}\gta$L$_*$) of the local luminosity function of galaxies even
assuming they evolve passively.

\section{The density of J-K$\ge3$ galaxies}
The co-moving spatial density of J-K$\ge3$ galaxies and its statistical 
uncertainty have been  estimated as
\begin{equation}
\rho=\sum_i{1\over V_{max}^i},~~~\sigma(\rho)=\Big[\sum_i\big({1\over V_{max}^i}\big)^2\Big]^{1/2}
\end{equation}
Thus, for each galaxy we computed the volume
\begin{equation}
V_{max}={\omega\over{4\pi}}\int_{z_1}^{z_{max}}{dV\over{dz}}dz
\end{equation}
where $\omega$ is
the solid angle subtended by the HDF-S area
and $z_{max}$ is the maximum redshift at which each galaxy
would be still included in the sample limited to Ks=22.
{ Since, for a given template the k-correction in Ks is nearly constant for $z>2-2.5$,
in the derivation of the $z_{max}$ of each galaxy we considered the average
 k-correction derived by the best fitting templates at redshift $z=z_{phot}$.}
The lower bound in the integration ($z_1$) should be defined by the 
J-K$>3$ color cut.
In Fig. 4 the evolution of the J-K color with redshift is
shown for different SFHs and two different values of the extinction.
It has been assumed that the star formation begins at $z_f=6$ with a Sal IMF.
   \begin{figure}
   \centering
   \includegraphics[width=9.5cm,height=8.5cm]{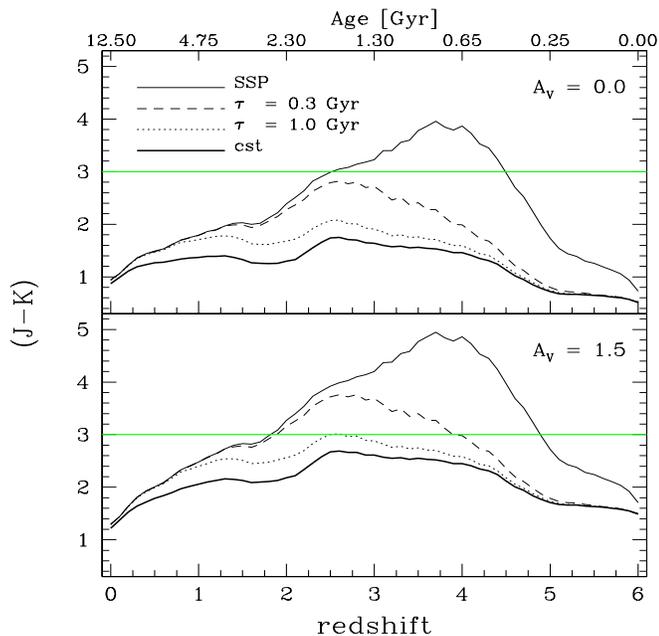}
\caption{Evolution of the J-K color as a function of redshift.
The different curves refer to the different SFHs quoted in the
 figure. 
All the models have been obtained with Sal IMF assuming a  star formation 
beginning at $z_f=6$ (Age=0) and two different values of
extinction (A$_V$=0 upper panel and A$_V$=1.5 lower panel).
The Age of the stellar population is shown on the top axis.
The horizontal line defines the J-K=3 color cut used to select
the three galaxies.
} 
\end{figure}
The figure clearly shows that the J-K$>3$ color cut does not define
a unique value of $z_1$ given the degeneracy with respect to the SFH 
and to the extinction. 
Thus, we assumed as $z_1$ the lower limit to the redshift of the
lowest redshift galaxy (HDFS-1269), i.e. $z_1=1.9$.
We find that at the average redshift $\langle z\rangle\simeq2.7$ the
co-moving density of galaxies brighter than Ks=22 with J-Ks$\ge3$ 
is $\rho=1.2\pm0.7\times 10^{-4}$ Mpc$^{-3}$.
In the derivation of $z_{max}$ we did not taken into account
the surface brightness dimming which would tend to lower the maximum
distance reachable.
Moreover, galaxies  with masses of about $10^{11}$M$_\odot$ at $z>2$
can escape the J-Ks$\gta3$ threshold   
due to the uncertainties in the flux measure and to intrinsically bluer colors.
{ Indeed, some of the galaxies selected by Labbe\`\ et al. (2003)
with color  J-K$>2.3$ at Ks$\le22$ turned out to have redshift
$z\gta2$ and masses of the order of $10^{11}$ M$_\odot$ 
(Fontana et al. 2003b).
Thus, it is likely that this density is an underestimate of the  
density of $\mathcal{M}_{stars}\gta10^{11}$M$_\odot$ galaxies at these redshifts.}

\section{Constraining the evolution of massive galaxies}
In this section we attempt to constrain the possible evolutionary
scenarios of the 3 massive galaxies.
In  \S 5.1 we concentrate on the formation and
evolution of the stellar populations, while in \S 5.2 we consider the 
evolution of the number density of massive galaxies.

\subsection{Constraining the star formation history}
Spectrophotometric models, as those adopted in the present work,
are not meant to describe the detailed evolutionary history
of  galaxies. 
Their aim is to describe the global observed photometric
properties of a galaxy  by means of an appropriate mix of stellar
populations and different SFHs.
Real galaxies are not necessarily expected to follow one of  
these SFHs.
Different combinations of them can provide good 
fits to the photometric data of a given galaxy  
implying different behavior of its past and future evolution 
(as clearly shown in \S 3).
Keeping these limits in mind, we have tried to 
constrain the SFH of the three massive galaxies.

In order to characterize the properties of their stellar populations 
we computed a more robust estimate of the age of the bulk of stars 
in the three EROs.
Indeed,  we cannot consider the age parameter  provided by the best-fitting models 
given the degeneracy of the SFHs.
{ We thus defined for each best fitting template
a mass weighted age of the stars.}
Any template, defined by a fixed age and SFH, can be seen as
the sum of SSPs with different ages.
Each SSP provides the fraction of the total mass which depends on its own
age and on the SFH describing the template itself.
We derived the mass weighted age by summing the ages of the SSPs,
each of them weighted on its mass fraction for each best-fitting template.
{ The age of the bulk of stars in each galaxy is defined by the range spanned by
the mass weighted age.}
Following this approach, we find that the bulk of the stars in
HDFS-1269, HDFS-822  and HDFS-850  have ages $1.7\pm0.5$ Gyr, 
$1\pm0.6$ Gyr and $0.6\pm0.3$ Gyr respectively.
Such estimates suggests that most of the stars have formed at $z_f\gta3.5$
in all three galaxies. 
The obvious question is  how the three galaxies have formed and/or assembled 
their stellar masses, i.e. how they evolved back in time.

{ Since in the hierarchical paradigm the age of the stars is not necessarily
correlated with the epoch of assembly of the objects (Moustakas and Somerville 2002), 
even our  estimate of the age of the bulk of stars 
does not place any constraint on the past assembly history (if any) of the three galaxies.
Thus, from the properties of the  stellar populations of the three galaxies
we cannot probe their possible formation through merging process.
Anyway, their redshift ($z\sim3$) place the possible merging event at
$z\gta3.5$ considering a dynamical time scale of $3\times10^8$ yr 
(Mihos \& Hernquist 1996).
On the other hand, we can probe the case in which the galaxies are 
the result of the SFHs resulting from the best-fitting procedure.
This procedure produced a set of acceptable SFHs only on the basis
of the photometry of the three galaxies. 
By means of the age of the stellar populations, the luminosity, the
stellar mass and the comoving density of the three galaxies, 
we now try to bound further this set of SFHs.
The following discussion and the results are, in fact, independent on 
the choice of the IMF.
In order to obtain the ages of the bulk of stars
estimated above in the three galaxies, the $\tau=0.1$ Gyr model 
(the lower bound of the SFHs) requires a formation redshift $z_f\simeq3.5$.
To produce a stellar mass of about 10$^{11}$ M$_\odot$ within 0.1 Gyr 
($3.3\lta z_f\lta3.5$), an average SFR$\simeq$1000 M$_\odot$ yr$^{-1}$ is needed
in this interval.
Such high SFR resembles that estimated by Ivison et al. (2002) for the 36 sources 
detected with SCUBA by Scott et al. (2002) and Fox et al. (2002) on
260 arcmin$^2$.
Given the comoving density of the three galaxies ($\rho=1.2\pm0.7\times 10^{-4}$ 
Mpc$^{-3}$) and the volume sampled by the HDF-S within $3.3\lta z_f\lta3.5$ 
($\sim3.1\times10^3$ Mpc$^3$) we expect $0.4\pm0.2$ of these progenitors 
over an area equal to that of the HDF-S.
This number would be consistent with the number of SCUBA sources expected 
on this area ($\sim0.7$) on the basis of the surface density measured by Scott et al.
Star formation time scales longer than 1 Gyr 
can  in principle account as well for the stellar masses already formed
and for the age of the bulk of stars, provided that $z_f>5$ and 
SFR$\le100$ M$_\odot$ yr$^{-1}$.
On the other hand, Fig. 4 suggests that a time scale  longer would not 
be consistent with the observed J-K color, unless to invoke an extinction in excess
to A$_V\simeq1.5-2$ mag.
However, even in this case, the star formation rate has to rapidly fade at 
$z\sim3$ to match the constraints imposed by the local 
density of  bright  galaxies.
Indeed, if the stellar mass already assembled ($\simeq10^{11}$ M$_\odot$)
by the three galaxies at $z_{phot}$ was lower than 50\% of the total 
mass in stars they could form from $z_f$ to $z=0$, they would be brighter 
than 2-3$L^*$ at $z=0$.
Thus, at $z=0$ we should measure a density of galaxies brighter than 2-3$L^*$
not lower than $1.2\times 10^{-4}$. 
The local density  of galaxies brighter than 2L$^*$ and 3L$^*$
is 1.8$\times 10^{-4}$ Mpc$^{-3}$ and 0.5$\times 10^{-4}$ Mpc$^{-3}$ respectively, 
i.e. comparable or lower than the density of the three galaxies.}
Thus, the photometry, the stellar masses already formed and assembled, the mean
ages of the stellar populations  together with
the constraints imposed by the local density of bright galaxies, 
suggest that the three galaxies are consistent with  
galaxies which at $z\sim3$ have formed most of their stellar mass
and which will follow a passive aging in time soon after few Myrs,
i.e. with a massive early type galaxy.

\subsection{Constraining the density evolution}
We have compared the density of local
massive galaxies with the one at redshift $z\simeq2.7$.
We have previously found that the  luminosity of the three galaxies 
at $z=0$ has to be  $\sim$L$^*$ in the case of passive evolution and
should not be brighter than $\sim$2L$^*$ in the case of a slightly longer 
star formation activity.
In this latter hypothesis, the three galaxies would represent $\sim$60\%
of the whole population of massive local galaxies (L$\ge$2L$^*$).
In the hypothesis of passive aging, we have to consider the local number 
density of early type galaxies.
By integrating the LF   of E/S0 galaxies brighter than L$^*$ 
($\mathcal{M}_{star}\simeq10^{11}$ M$_\odot$) we obtained a density of
 3$\times10^{-4}$ Mpc$^{-3}$,  
where we used $\phi^*_{E/S0}=1.5\times10^{-3}$ Mpc$^{-3}$ (Marzke et al. 1998).
Thus, the three galaxies account for about 40\% 
of the local population of early type galaxies.
This result implies also that $\sim$40\% of the stellar mass content of local 
early type galaxies was already in place at $z\gta2.5$ and that
their number density cannot decrease by more than a factor 2.5-3 from $z=0$ 
to $z\simeq3$.
{ In the semi-analytical rendition of hierarchical models by Kauffmann  
and Charlot (1998a,b), the predicted number density of 
$\mathcal{M}_{star}\ge10^{11}$ M$_\odot$ ellipticals decreases by a
factor $\sim8$.
Indeed, the predicted density of $\mathcal{M}_{star}\ge10^{11}$ M$_\odot$ 
ellipticals at $z\simeq2.5$ is about $5-6\times10^{-5}$ Mpc$^{-3}$,  
a factor 2 lower than our estimate.}
The same conclusion is reached considering the predictions of Moustakas and 
Somerville (2002). They  predict a number density of halos hosting  local bright
(L$>$L$^*$) elliptical galaxies (gEs) at $z\simeq2.5$  more than a factor two 
lower than our estimate.
The semi-analytic models of galaxy formation in the hierarchical clustering
scenario by Baugh et al. (1998, 2002) predict, in fact, no such massive 
galaxies at redshift $z>2$.
These results are summarized in Fig. 5.
The figure suggests that the evolution of the number density of massive 
($\mathcal{M}_{stars}\simeq10^{11}$ M$_\odot$) galaxies with redshift 
is slower than that predicted by the current hierarchical models, 
at least in the redshift range $0<z<3$.
It is worth noting that, by assuming a  value of 
$\mathcal{M}_{star}\simeq10^{11}$ M$_\odot$ 
for the three EROs, we derive at $z\simeq2.7$ a stellar mass density 
$\rho_{star}=1.2(\pm0.7)\times10^7$ M$_\odot$ Mpc$^{-3}$.
This density is  $\sim30\%$ of the total stellar mass  
at this redshift (Fontana et al. 2003b).

The density we estimated is lower than the density of ellipticals 
(2.16$\pm0.6\times10^{-4}$ Mpc$^{-3}$) spectroscopically 
identified at $0.85<z<1.3$ by Cimatti et al. (2002).
On the other hand,  this estimate  includes mainly
ellipticals less massive than $10^{11}$ M$_\odot$.
{ Indeed, given the limiting magnitude  K=19.2 of their spectroscopic sample,
they are sampling the counterpart of local  L$\gta0.3$L$^*$ 
($\mathcal{M}_{star}>10^{10}$ M$_\odot$) ellipticals 
in that redshift range. 
Our estimate is also lower  than  the density of 
$\mathcal{M}_{star}\ge10^{11}$ M$_\odot$ galaxies at $z\sim1$
estimated by Drory et al. (2001).
This is expected since they use an approach which maximizes the stellar
mass of the galaxies for any K-band luminosity at any redshift. 
Thus their estimate represent the upper limit of $\mathcal{M}_{star}\ge10^{11}$ 
M$_\odot$ galaxies at that $z$. }

   \begin{figure}
   \centering
   \includegraphics[width=9cm]{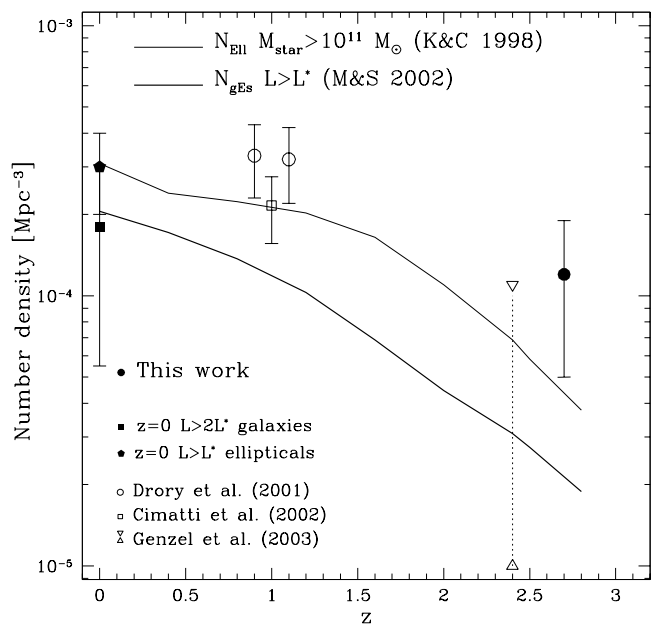}
\caption{Number density of galaxies as a function of redshift. 
The filled circle is our estimate of
the density of $\mathcal{M}_{stars}\simeq10^{11}$ M$_\odot$ 
galaxies at $z\simeq2.7$.
The filled triangle
is the local number density of L$>2L^*$  galaxies 
obtained by integrating the LF of Cole et al. (2001; see \S 4). 
The filled pentagon is the density of local  L$>L^*$ early type galaxies
derived by the LF of Marzke et al. (1998).
The open square is the density of L$\gta0.3L^*$ ellipticals from Cimatti et al. (2002) 
and the open circles are the upper limit to the density 
of $\mathcal{M}_{stars}\ge10^{11}$ M$_\odot$ galaxies from Drory et al. (2001)
(see text for details).
The two open triangles connected by the dotted line represent the range of values 
of the density of $\mathcal{M}_{stars}>10^{11}$ M$_\odot$ galaxies as 
derived by Genzel et al. (2003) from the SCUBA lens survey. 
The thin line
is the density evolution of ellipticals with stellar mass 
greater than $\mathcal{M}_{stars}=10^{11}$ M$_\odot$ from Kauffmann and
Charlot (1998a,b; K\&C 1998); the thick line is 
the density evolution of dark matter halos hosting present-day bright (L$>$L$^*$)
ellipticals (gEs) from Moustakas and Somerville (2002; M\&S 2002). 
} 
\end{figure}

\section{Discussion and Conclusions}
We have presented the analysis of the three galaxies selected on the HDF-S 
on the basis of their unusually red  (J-K$\ge3$) color.
We have constrained their redshifts and  estimated their stellar mass 
content by comparing the photometric data with population synthesis
models obtained for different IMFs, SFHs, metallicity and extinction values.
The grid of templates we used includes a set of template  derived from   
the spectral models of Charlot \& Longhetti (2001) which take into 
account consistently  the emission from stars  and from the gas.
This has allowed us to exclude the possibility that the 
observed extreme colors are dominated by emission lines.
We find that  one of the galaxies (HDFS-1269) is at $z_{phot}\simeq2.4$
(P($\chi^2$)=0.92)
while the other two (HDFS-822 and HDFS-850) are at $z_{phot}\simeq3$ 
(P($\chi^2$)$\ge0.98$).
All three galaxies have already assembled a stellar mass
of about $10^{11}$ M$_{\odot}$ at the relevant redshift. 
The fact that the two galaxies at $z\sim3$ have assembled this mass
places the possible merging event of their formation at  z$\gta3.5$ considering 
a dynamical time scale of 3$\times10^8$ yr 
(e.g. Mihos \& Hernquist 1996).
The inferred   mass weighted age of the stellar populations 
places the formation of their bulk  at $z_f>3.5$ in all 
three galaxies suggesting a substantial amount of star formation  
at these redshifts  in addition to the one derived by LBGs  
(e.g. Ferguson et al. 2002).
Galaxies with stellar masses $\mathcal{M}_{star}\gta10^{11}$ M$_{\odot}$ fully
 assembled at $z>2$ were previously found by other authors.
For instance, Francis et al. (2001) find two luminous extremely red galaxies 
at $z\sim2.4$ whose radial profiles suggest they are elliptical galaxies.
They estimate a mass $\sim10^{11}$ M$_{\odot}$ and an age of 
$\sim7\times10^8$ yr for their stellar populations. 
Shapley et al. (2001) find some galaxies with such high stellar masses 
and old ages among their  LBGs at $z\simeq3$.
Genzel et al. (2003) estimate a conservative lower limit of $1.4\times10^{11}$
M$_{\odot}$ to the stellar mass of the sub-mm source at $z\simeq2.8$ they studied. 
Contrary to these and our findings,  Dickinson et al. (2002) 
do not find galaxies more massive than $10^{11}$ M$_{\odot}$ at $z>2$ 
in the HDFN. 
The small area  covered by the HDFs could
be the reason of this discrepancy even if surface densities of J-K$\gta3$ 
galaxies comparable or higher than that in the HDF-S are found in the other 
pencil beams reaching similar depth (e.g. Bershady et al. 1998; 
Totani et al. 2001).

We found that the three 
galaxies must have already formed most of their stellar mass and 
that they cannot follow an evolution significantly different from a passive aging.
This suggests that J-K$>3$ galaxies are most likely  in the post starburst 
phase rather than  in the starburst phase of their formation as hypothesized
by Totani et al. (2001).
These findings strongly support the thesis  that J-K$>3$ galaxies are the high-z 
counterpart of local $\mathcal{M}_{star}\gta10^{11}$ M$_{\odot}$
early type galaxies and agree with the recent finding
of an increasing clustering of high-z galaxies with redder colors (Daddi et al. 2003;
Roche et al. 2003).
We estimated a co-moving density of galaxies brighter than Ks=22 and redder than J-K$=3$ 
 $\rho=(1.2\pm0.7)\times 10^{-4}$ Mpc$^{-3}$ at the average redshift 
$\langle z\rangle\simeq2.7$ which should be considered an underestimate 
of the number density of galaxies with $\mathcal{M}_{star}\gta10^{11}$ M$_{\odot}$ 
at that $z$.
This value is about a factor two higher than the predictions of  
hierarchical models renditions by Kauffmann \& Charlot (1998a,b) and Moustakas \&
Somerville (2002).
In the hypothesis of passive evolution,  their luminosities 
at $z=0$ would be L$\ge$L$^*$.
By comparing the density of local L$\ge$L$^*$ early type galaxies
with our estimate  we find that their density cannot decrease by  more than a factor 
2.5-3 from $z=0$ to $z\simeq3$ suggesting that up to 40\% of the stellar mass 
contained  in local massive galaxies was already in place at $z\gta2.5$.

\begin{acknowledgements}
\thanks{This work is based on observations made with the ESO-VLT telescopes at 
Paranal Observatory under program ID 164.O-0612 and with the NASA/ESA Hubble Space 
Telescope, obtained from the data archive at the Space Telescope Institute. 
STScI is operated by the association of Universities for Research in Astronomy, 
Inc. under the NASA contract  NAS 5-26555.} 
\end{acknowledgements}

\end{document}